%% file: AnalRepSSVPlate.tex
\documentclass[11pt,twocolumn,twoside]{article}
\usepackage{graphicx} 
\usepackage{amsmath}  
\usepackage{txfonts}
\usepackage{bm}
\input{jasnaoe-e.tex}
\input{deffile-e.tex}

\begin{document}
%
\title{\fontsize{16.5pt}{0pt}{\bf 
The Analytical Representations for 2-D Flows around a Semi-Submerged Vertical Plate in a Uniform Stream$^\dagger$
}\\[-1.5mm]{\normalsize 
}}%
%
\eauthor{
\hspace*{52mm} {by} \quad
\ename{Katsuo Suzuki$^*$}{}     \hspace{6.7mm} %
\ename{Shigeyuki Hibi$^{**}$}{}     
}
\abst{
The complex potentials representing flows around a vertical plate semi-submerged in a uniform stream are derived in analytical forms by the reduction method.   
They are composed from the regular solution and a weak singular eigen solution.   
The linear combinations of them represent some flows such as regular flow, zero-vertical flux flow, flow satisfying Kutta condition and wave-free flow.   
The wave resistances of the flows are also obtained in analytical forms.   
The analytical solution obtained by Bessho-Mizuno(1962) has a possibility that it does not satisfy the boundary condition on the plate.
}
\maketitle
%
\begin{table}[b]
\footnotesize 
\hrulefill \\
\par\vspace*{-2mm}
\begin{tabular}{ll}
*    & \hspace*{-2mm}Professor Emeritus of National Defense Academy  \\[0.5mm]
**   & \hspace*{-2mm}National Defense Academy   \\[0.5mm]
\multicolumn{2}{l}{$\dagger$ Translated from Journal of Japan Society of Naval Architects and Ocean Engineers, Vol. 17,  pp.1-8, 2013} 
\end{tabular}
\vspace*{-2mm}
\end{table}
%
%
\footnotesize\baselineskip\mybaselineskip

\jsection{Introduction}
The purpose of this thesis is to find an analytical solution expressing the flow around the vertical flat plate semi-submerged in a two dimensional uniform stream.

The water surface condition is assumed to be linear.

As for the analytical solution in the two-dimensional linear water wave problem, a solution by Ursell\cite{U} about a flow around the semi-submerged plate placed in a periodic incident wave is famous.
Ursell found analytical solution expressing local wave in the form of integral including Bessel function and succeeded in obtaining the reflection and transmission coefficients by the flat plate using modified Bessel function.

On the other hand, an analytical solution on the problem of the present thesis was reported by Bessho-Mizuno\cite{BM}.
This solution is a form of analytical expression for the solution of the integral equation derived from doublet distribution.

It was stated in Bessho-Mizuno's paper that the function used there may be expressed by the Bessel function of the imaginary variable.
However, it is not a easy form in comparison with Ursell's solution.
Moreover, as described in Chapter 3 of this paper, there is a doubt as to if the boundary condition is correctly satisfied.

In this paper we apply the reduction method\cite{W} to this problem and obtain the same type of solution as Ursell's.
This solution is similar to the method of Bessho-Mizuno in using the reduction method.
However,  paying attention to the fact that the solution of the reduction method is not unique, we succeeded in getting solution satisfying the boundary condition correctly.
In this thesis we call this solution a regular solution.
In the regular solution, the streamline passing through the position corresponding to the reciprocal of the wavenumber on the infinite upstream water surface reaches the flat plate surface.

Next, we showed that there exists further a solution satisfying the homogeneous boundary condition on the flat plate, if we allow weak singularity at the intersection of the flat plate and the static water surface.
We also showed that various flows can be created by varying the strength of the sigularity of the homogeneous singular solution.
It is possible to create a flow which is called a zero-vertical-flux flow in which a streamline passing through the upstream water surface reaches the flat plate surface.
Furthermore, it is possible to make a solution satisfying Kutta condition at the lower end of the flat plate, and even to create a wave-free flow which is symmetrical with respect to the up- and down-stream of the flat plate and generates no subsequent trailing waves.

Finally, by comparing the present solution with a solution by the boundary element method, we discuss the possibility that solution by Bessho-Mizuno\cite {BM} might not satisfy the boundary condition on the flat plate.

In this paper, we only discuss the representations and the properties of the solutions in a boundary value problem from a pure mathematical point of view.
We are not considering interesting matters such as wave resistance and momentum loss phenomenon due to wave breaking or eddy making etc. to which the singular homogeneous solution is related. 

\jsection{Reduction method}
%
\jsubsection{Regular solution}

\begin{figure}[bh]
        \vspace*{1mm}
        \begin{center}
               \includegraphics[width=0.60\linewidth,angle=90,bb=100 180 500 670,clip]
                        {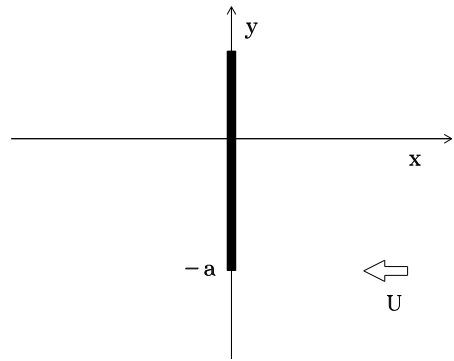}
                \caption{Co-ordinate system. }                                                   \label{fig201}
        \end{center}
        \par\vspace*{-3mm}
\end{figure}

We take the coordinate system as shown in Fig.\ref {fig201}.

Uniform flow is assumed to flow in the negative direction of the $x$ and the depth of the vertical flat plate is $a$.

The flow field is represented by a complex potential.
\begin{equation} 
        \left.\begin{aligned}
                F(z)=&-Uz+w(z)  \\
                    =&\,\gPhi(x,y)+i\gPsi(x,y),      \\
                z=&x+iy.                                                         \\
        \end{aligned}\hspace{0.5cm} \right\}                                    \label{equ201}
\end{equation}

\clearpage
The linear free surface condition can be written as (see Eq.(20.3')\cite{W})
\begin{equation}  
        \left.\begin{aligned}
                &\mathtt{Re}\{L[w(z)]\}=0\ \ \ \     on\ \ \  y=0,      \\
                &\hspace{8mm}L\equiv\frac{d}{dz}+i\nu-\mu,              \\
                &\hspace{8mm}\nu=g/U^2,                             \\
        \end{aligned}\hspace{0.5cm} \right\}                                    \label{equ202}
\end{equation}

where $\nu$ is the wave number and $\mu$ is the virtual friction coefficient ($\mu\to +0$).

The boundary condition on the flat plate can be written in the following two forms.

If we write with velocity
\begin{equation}  
          \mathtt{Re}\{\frac{dw}{dz}(z)\}=U\ \ \ on\ \ \ x=0,\ -a\leqq y\leqq 0.                                                                                                                                \label{equ203}
\end{equation}
If expressed by a stream function
\begin{equation}  
          \gPsi(0,y)=\gPsi_H\ (Real\, Const.)\ \ \ on\ \ \ x=0,\ -a\leqq y\leqq 0.                                                                                                                                               \label{equ204}
\end{equation}
The purpose of this thesis is to obtain an analytical representation of the perturbed complex potential $w(z)$ under the above boundary conditions.

We use the reduction method\cite{W} to find the solution.
If the computation of the free surface condition is applied to the perturbed complex potential $w(z)$, it becomes a reduced function like so-called local wave which does not have a wave term.
First, we solve analytically the boundary condition that this function should satisfy, and finally apply the inverse operator $L^{-1}$ to obtain the original solution.
This method is similar to that used by Russian scholars\cite{W}

First, we apply the operator $L$ to $w(z)$ to get a reduced function.
\begin{align}  
        Uf(z)=&L[w(z)]  \notag \\
             =&\frac{dw}{dz}(z)+i\nu w(z).                                              \label{equ205}
\end{align}
The free surface condition becomes simple as
\begin{equation} 
        \mathtt{Re}\{f(z) \}=0\ \ \ on\ \ \ y=0.                                \label{equ206}
\end{equation}

On the flat plate, we obtain the following condition using both boundary conditions (\ref{equ203}) and (\ref{equ204}).
\begin{equation} 
         \mathtt{Re}\{f(z) \}=1-\frac{\nu}{U}\,\gPsi_H-\nu y\ \ \ on\ \ \ x=0, \ -a\leqq y\leqq 0.                                                                                       \label{equ207}
\end{equation}

It should be noted here that this condition is merely a necessary condition since the expression (\ref{equ207}) uses the two equivalent boundary conditions.
It is necessary to return to the condition (\ref {equ203}) or (\ref{equ204}) later.

Also we write the expression (\ref{equ205}) in the following form.
\begin{equation} 
         \frac{dw}{dz}(z)=-i\nu w(z)+Uf(z).                                     \label{equ208}
\end{equation}
From this equation we can see that the singularity of the derivative (complex velocity) of $ w (z) $ is identical to the singularity of $ f (z) $.

Next, we notice that if the (\ref {equ206}) and (\ref{equ207}) expressions are continuous at the top of the plate ($ x = y = 0 $), we have.
\begin{equation} 
        \gPsi_H=U/\nu.                                                                          \label{equ209}
\end{equation}
We call the solution $ w _ R (z) $ which satisfies this condition as a regular solution, and we will find this solution first. 
This condition means that the streamline flowing at the position of $1/\nu$ under the water surface of the infinite upstream reaches the surface of the flat plate.
Then, the condition (\ref{equ207}) is given by
\begin{equation} 
         \mathtt{Re}\{f(z) \}=-\nu y\ \ \ on\ \ \ x=0, \ -a\leqq y\leqq 0.                                                                                                                                      \label{equ210}
\end{equation}

Now we know that a regular solution satisfying the condition (\ref {equ206}) and equation (\ref {equ210}) can be given by the following expression from the knowledge of theory of complex function\cite{Mu}.
\begin{equation}
        \left.\begin{aligned}
                f(z)=\nu&a\,[F_1(z)+C_1f_0(z)], \\
                F_1(z)=&\,i\,\frac{z}{a}\,\biggl(1-\sqrt{1+\frac{a^2}{z^2}}\ \biggr),\\
                f_0(z)=&\,i\,\frac{a}{z}\,\frac{1}{\sqrt{1+a^2/z^2}}.
        \end{aligned}\hspace{0.5cm} \right\}                                    \label{equ211}
\end{equation}
The branch line of the root is assumed to be $ x = 0, \ | y | <a $.
Here, $ f_0 (z) $ satisfies a homogeneous condition for the condition (\ref {equ210}), so $ C_1 $ can be an arbitrary real constant.
This constant needs to be determined later so as to satisfy the original boundary condition (\ref {equ203}) or (\ref {equ204}) as described above.
$ F_1 (z) $ is regular all over the plane including the origin.
$f_0 (z) $ is regular on the lower half plane including the origin except that it has the singularity $ 1 / \sqrt {z + ai} $ of the complex velocity at the edge of the flat plate ($ z = -ai $) .
In Ursell's problem, fortunately there exists only an indefinite real constant instead of this homogeneous solution and the problem becomes simplified.

\begin{figure}[bp]
        \vspace*{1mm}
        \begin{center}
                \includegraphics[width=0.90\linewidth,angle=0,bb=1 1 360 280,clip]
                        {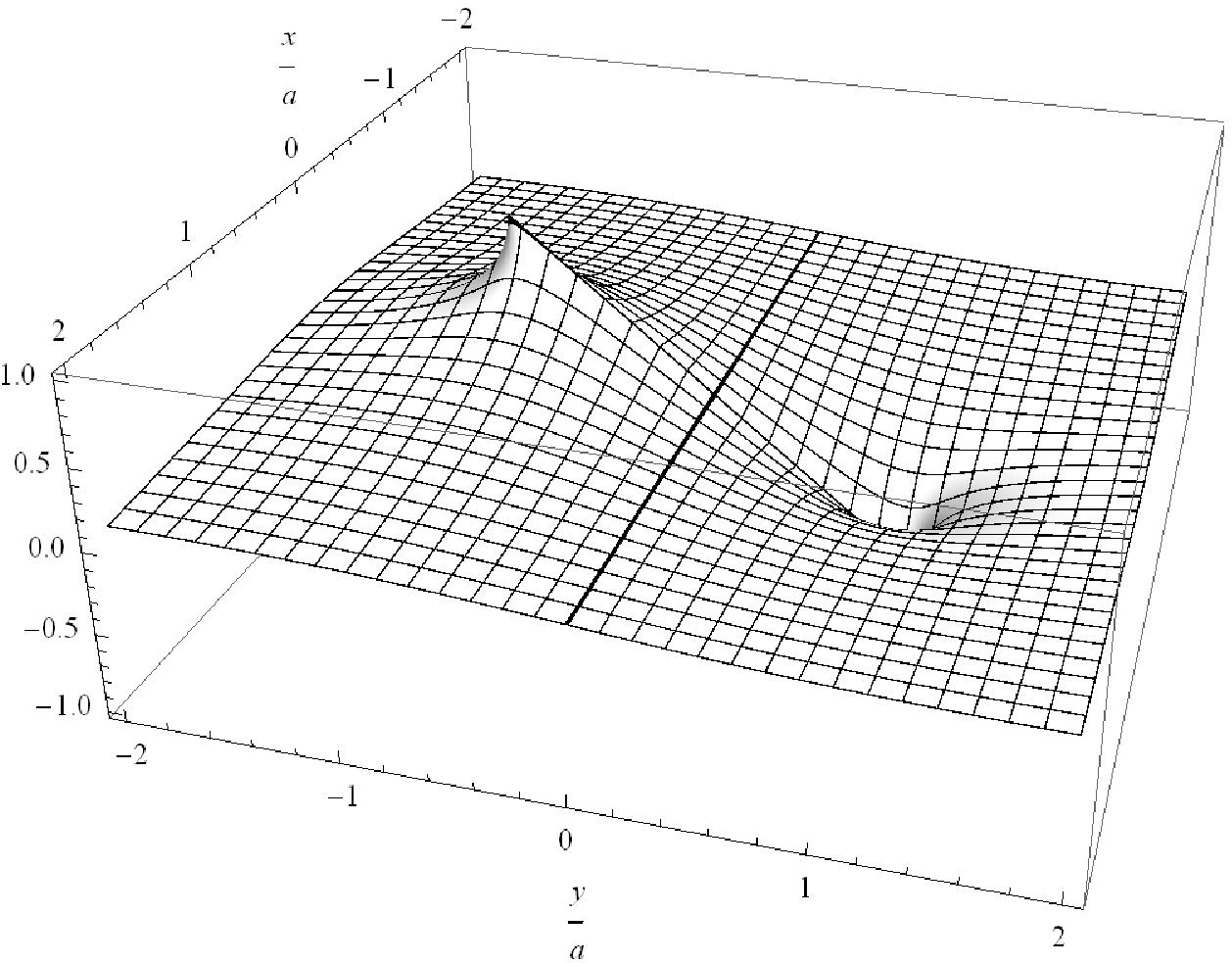}
                \caption{3-D plot of $\mathtt{Re}\{F_1(z)\}$. }                 \label{fig202}
        \end{center}
        \par\vspace*{-3mm}
\end{figure}

\begin{figure}[tp]
        \vspace*{1mm}
        \begin{center}
                \includegraphics[width=0.90\linewidth,angle=0,bb=1 1 360 280,clip]
                        {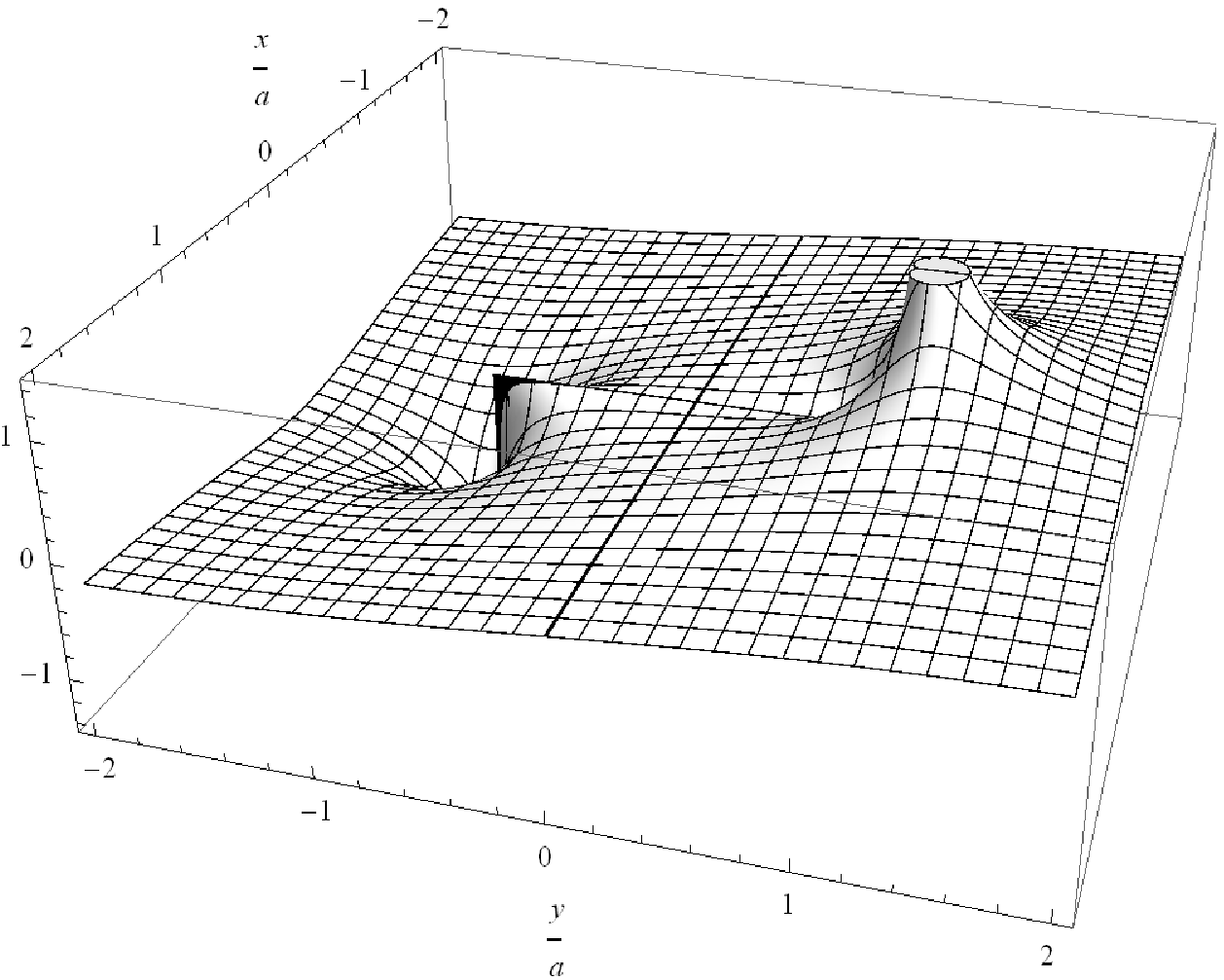}
                \caption{3-D plot of $\mathtt{Re}\{f_0(z)\}$.}                  \label{fig203}
        \end{center}
        \par\vspace*{-3mm}
\end{figure}

Fig.\ref {fig202} plots the real part of $ F_1 (z) $ three dimensionally.
We can see that the value takes $ 0$ on the $ x $ axis and $ -y $ in the region of $-a<y<a$ on $y $ axis. 
Fig.\ref {fig203} shows a three-dimensional plot of the real part of the $ f_0 (z) $ function.
The values are $0$ on the $x$ axis and in $-a<y<a$ on the $y$ axis.

We express the above functions in integral form including Bessel functions to make the $ L^{-1} $ operator work on them.
We also introduce $ f_2 (z) $.
\begin{align} 
        F_1(z)=&\mp i\int_0^\infty \frac{1}{k}J_1(ak)e^{\mp kz}dk   \notag\\
              =&\mp i\,\frac{a}{2}\int_0^\infty  \{J_0(ak)+J_2(ak)\}e^{\mp kz}dk, 
                                                            \label{equ212}
\end{align}
\begin{equation} 
        \hspace{-17mm}f_0(z)=\pm\,i\,a\int_0^\infty J_0(ak)e^{\mp kz}dk,                                                                                                                                         \label{equ213}
\end{equation}
\begin{equation} 
        \hspace{-17mm}f_2(z)=\pm\,i\,a\int_0^\infty J_2(ak)e^{\mp kz}dk.                                                                                                                                         \label{equ214}
\end{equation}
These are Laplace transformations of Bessel functions.
The double-sign corresponds to $ x \gtrless 0 $.
The expression (\ref {equ212}) becomes the following from the definition.
\begin{equation} 
        F_1(z)=-\frac{1}{2}[f_0(z)+f_2(z)].                             \label{equ215}
\end{equation}
By the way, there is the following relationship.
This function appears in Ursell's method.
\begin{align} 
        f_1(z)=&a\frac{d}{dz}F_1(z)  \notag \\
              =&i\,a\int_0^\infty J_1(ak)e^{\mp kz}dk.                          \label{equ216}
\end{align}

Similarly as the relation of equation (\ref{equ205}?jthe complex potentials corresponding to $F_1(z),\,f_0(z)\,$ and $\,f_2(z)$ are given by $W_1(z),\,w_0(z)\,$ and $\,w_2(z)$, respectively.
\begin{equation}
        \left.\begin{aligned}
                UF_1(z)=&L[W_1(z)],     \\
                Uf_0(z)=&L[w_0(z)],     \\
                Uf_2(z)=&L[w_2(z)],     
        \end{aligned}\hspace{0.5cm} \right\}                                    \label{equ217}
\end{equation}
\begin{equation} 
        W_1(z)=-\frac{1}{2}[w_0(z)+w_2(z)].                                     \label{equ218}
\end{equation}
Then, regular solution $ w_R (z) $ can be written as follows.
\begin{equation}
        \left.\begin{aligned}
                w_R(z)=&\nu a[W_1(z)+C_1w_0(z)] \\
                      =&-\frac{1}{2}\nu a[w_2(z)+C'_1w_0(z)],   \\
                &C'_1=1-2C_1.   \\
        \end{aligned}\hspace{0.5cm} \right\}                                    \label{equ219}
\end{equation}
If $ w_0 (z), w_2 (z) $ can be obtained, the solution $ w_R (z) $ can be found immediately from the above equation.

Let's first find $ w_0 (z) $ from $ f_0 (z) $.
We transform $ f_0 (z) $ 
keeping in mind that we are thinking in the area of $ y <0 $.
We move the integration path onto the positive and negative imaginary axis corresponding to the positive and negative of $ x $.
\begin{equation} 
        f_0(z)=\pm\,i\,a\int_0^{\mp i\,\infty} J_0(ak)e^{\mp kz}dk.   \label{equ220}
\end{equation}
Here, if we convert the variable $ k = \pm it $, we get:
\begin{equation} 
        f_0(z)=-a\int_0^{\infty} I_0(at)e^{-itz}dt.                     \label{equ221}
\end{equation}
Now, we are ready to apply $ L^{-1} $.
\begin{align} 
        w_0(z)=&L^{-1}\bigl[Uf_0(z)\bigr]  \notag \\
        =&-iaU\int_0^\infty\frac{1}{t-\nu-i\mu}\,I_0(at)e^{-itz}dt,  \label{equ222}
\end{align}
where $I_0$ and $I_2$ below are modified Bessel functions.
Again, we move the integration path on the negative and positive imaginary axis according to positive and negative of the sign of $ x $.
Then, it is necessary to note that when $ x <0 $, a residue term occurs at $ t = \nu $. 
We decompose $ w_0 (z) $ into the sum of the local wave component $ w_{0_L} (z) $ and the free wave component $ w_{0_F} (z) $.
\begin{equation} 
        w_0(z)=w_{0_L}(z)+w_{0_F}(z).                                                    \label{equ223}
\end{equation}
Furthermore, if we convert the variable $ t = \mp ik $ and return it to the original variable, we get the following local wave component.
\begin{equation} 
        w_{0_L}(z)=-iaU\int_0^\infty\frac{1}{k\mp i\nu}J_0(ak)e^{\mp kz}dk.                                                                                                                                     \label{equ224}
\end{equation}
The free wave term obtained from the residue term is as follows.
\begin{equation} 
        w_{0_F}(z)=
                \begin{cases}
                        \ \ \ 0 &for\ \ \ x>0,                          \\
                        2\pi aUI_0(\nu a)\,e^{-i\nu z} &for\ \ \ x<0.
                \end{cases}                                                                             \label{equ225}
\end{equation}

We can find the $ w_2 (z) $ function by performing the same operation.
\begin{equation} 
        w_2(z)=w_{2_L}(z)+w_{2_F}(z),                                           \label{equ226}
\end{equation}
\begin{equation} 
        w_{2_L}(z)=-iaU\int_0^\infty\frac{1}{k\mp i\nu}J_2(ak)e^{\mp kz}dk,                                                                                                                                     \label{equ227}
\end{equation}
\begin{equation} 
        w_{2_F}(z)=
                \begin{cases}
                        \ \ \ 0 \hspace{25mm}\,&for\ \ \ x>0,           \\
                        -2\pi aUI_2(\nu a)\,e^{-i\nu z}&for\ \ \ x<0.
                \end{cases}                                                                     \label{equ228}
\end{equation}

We decompose the integrand of the expression (\ref {equ224}), (\ref{equ227}) into the real and the imaginary parts as follows.
\begin{align} 
         \frac{1}{k\mp i\nu}e^{\mp kz}=\frac{1}{k^2+\nu^2}
                \bigl\{[&k\cos ky+\nu\sin ky]                  \notag \\
  \pm i[&\nu\cos ky-k\sin ky]\bigr\}e^{\mp kx}.                         \label{equ229}
\end{align}
The form of the above expression would be convenient to integrate the function (\ref {equ224}), (\ref{equ227}) numerically.
In the numerical example shown in this thesis, the above expression is used, and, by the adaptive numerical integration method (Quanc8)\cite{F}, the direct numerical integration is almost successful except in the neighborhood of $ x = 0 $.
For $ x = 0 $, the some analytical forms are shown in the appendix in the form obtained by further decomposing the above formula.

Since the form of the solution has been determined above, we must return again to the boundary condition (\ref {equ203}) or (\ref{equ204}) on the flat plate.
In this case, it is convenient to use a stream function, that is, an expression (\ref {equ204}) in which the value of the equation (\ref {equ209}) is substituted.
That is, at $ x = 0 $ the expression (\ref{equ219}) must satisfy the following.
\begin{equation} 
-\frac{1}{2}\,\nu a\,\mathtt{Im}\{w_2(iy)+C'_1w_0(iy) \}=U(y+1/\nu).                                                                                                                                    \label{equ230}
\end{equation}
According to the appendix, we use the modified Bessel function at $ x = 0, -a \leqq y \leqq 0 $ to obtain
\begin{equation} 
       \frac{1}{Ua}\mathtt{Im}\{w_0(iy)\}=-K_0(\nu a)\,e^{\nu y},\label{equ231}
\end{equation}
\begin{equation} 
        \frac{1}{Ua}\mathtt{Im}\{w_2(iy)\}=-\frac{2}{\nu a}\bigl(\frac{1}{\nu a}+\frac{y}{a}\bigr)+K_2(\nu a)\,e^{\nu y}.               \vspace{2mm}                    \label{equ231?f}
\end{equation} 
From the above, we obtain the following result.
\begin{equation} 
         C'_1=\frac{K_2(\nu a)}{K_0(\nu a)}.                                    \label{equ232}
\end{equation}
We substitute the expressions (\ref {equ223}-\ref {equ225}) and (\ref {equ226}-\ref {equ228}) into the equation (\ref {equ219}) and give the constant by (\ref{equ232}). 
Then, analytical representation of the regular solution $ w_R (z) $ is obtained.

In the flow represented by the regular solution, the wave resistance acting on the flat plate can be analytically obtained as follows.
The free wave is represented as follows.
\begin{equation}
        \left.\begin{aligned}
                \eta_F(x)=&\frac{1}{U}\mathtt{Im}\{w_F(x)\}     \\
                         =&A_R\sin\nu x,        \\
                A_R/a=-\pi\nu a&\bigl[\frac{K_2(\nu a)}{K_0(\nu a)}I_0(\nu a)-I_2(\nu a)\bigr].
        \end{aligned}\hspace{0.5cm} \right\}                                     \label{equ233}
\end{equation}
Therefore, the wave resistance $ R_W $ can be expressed by
\begin{align} 
        C_W=&\frac{R_W}{\rho ga^2}  \notag \\
        =&\frac{1}{4}(A_R/a)^2.                                                                 \label{equ234}
\end{align}

\begin{figure}[tp]
\vspace*{1mm}
\begin{center}
\includegraphics[width=0.65\linewidth,angle=90,bb=352 285 1489 2100,clip]{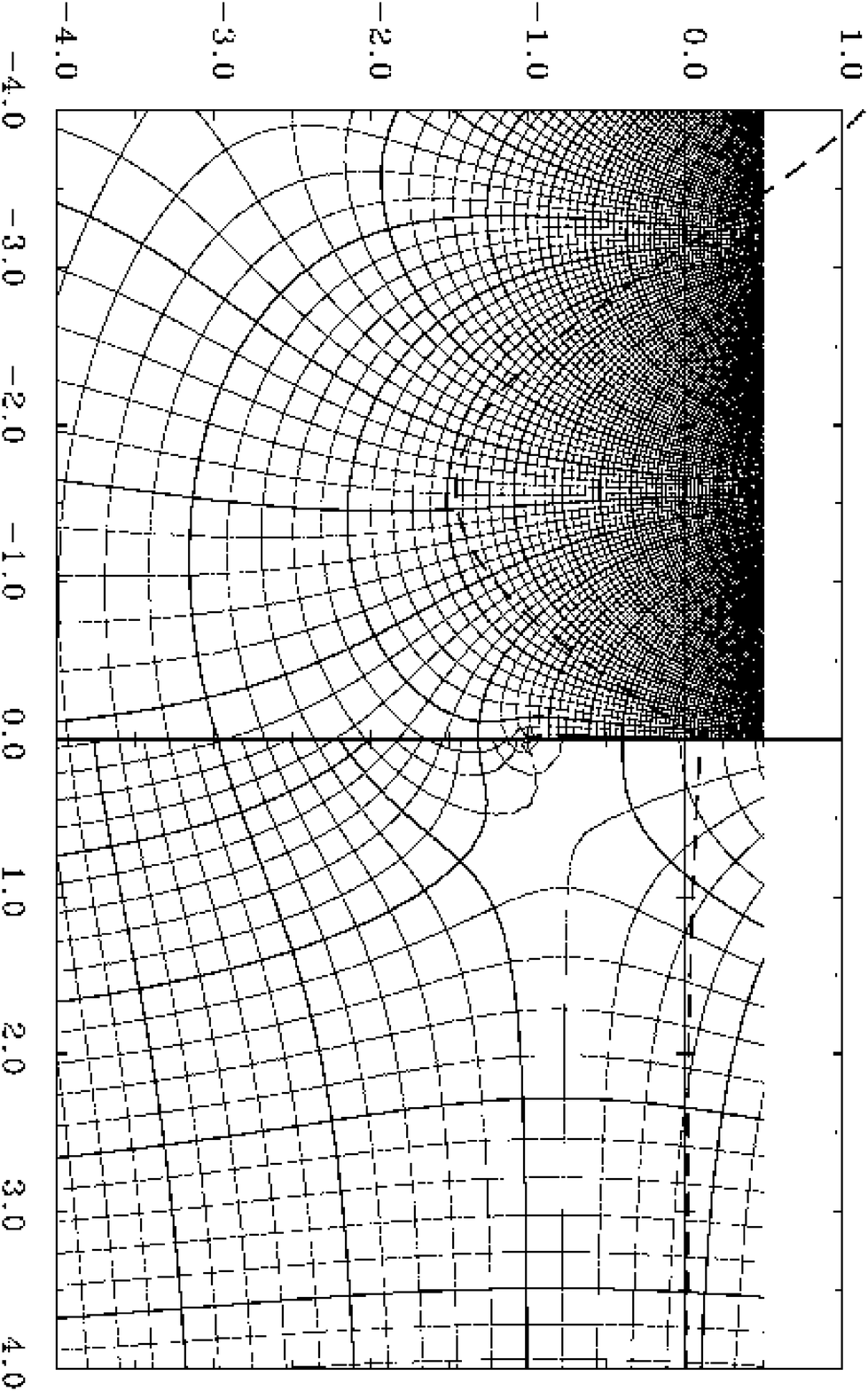}
\caption{Streamlines and equi-potential lines for regular solution ($\nu a=1.0,\ \gPsi_H/Ua=1.0,\ \gDelta\gPsi/Ua=0.2$). }                       \label{fig204}
\end{center}
\par\vspace*{-3mm}
\vspace*{1mm}
\begin{center}
\includegraphics[width=0.65\linewidth,angle=90,bb=352 285 1489 2100,clip]{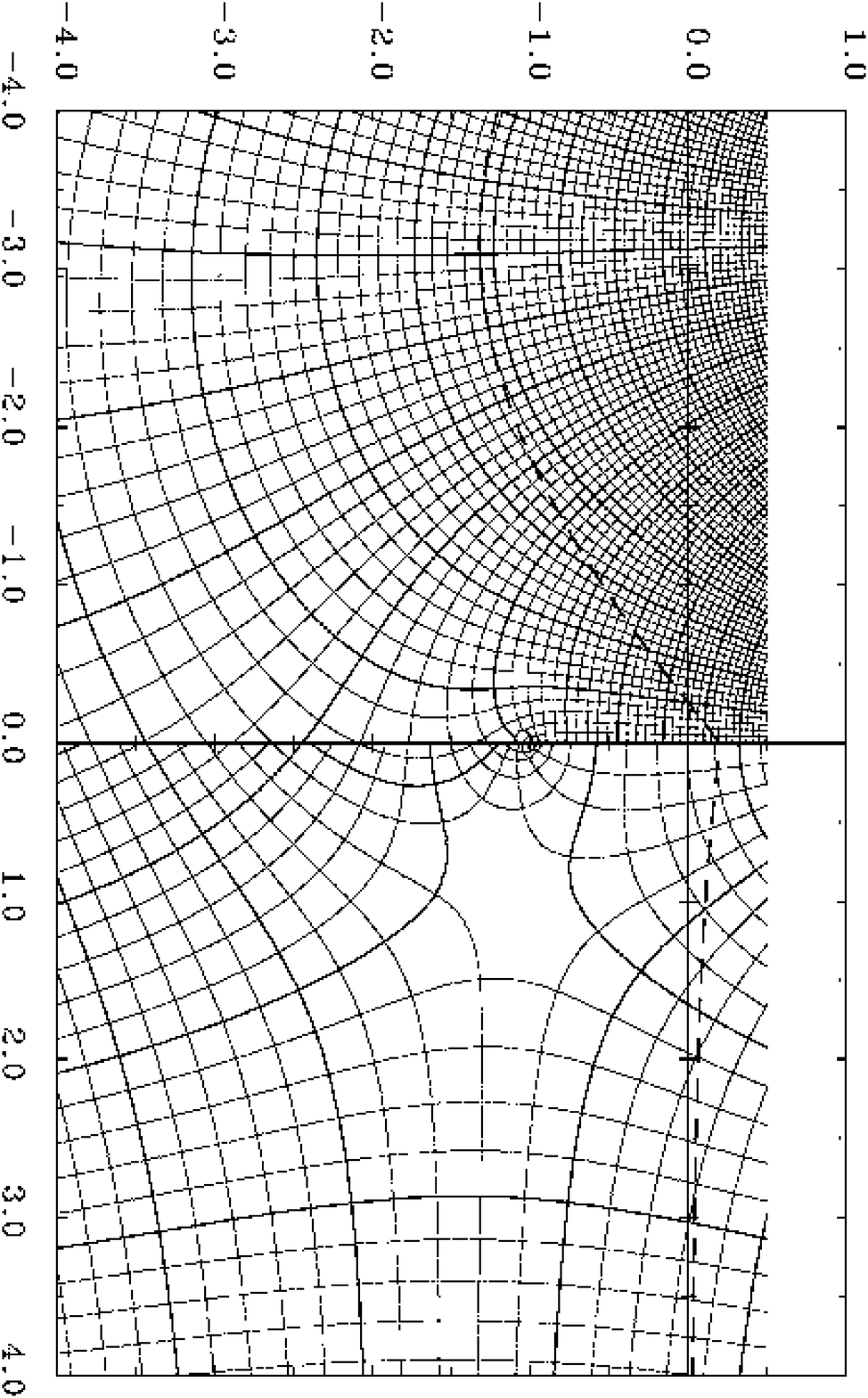}
\caption{Streamlines and equi-potential lines for regular solution ($\nu a=0.5,\ \gPsi_H/Ua=2.0,\ \gDelta\gPsi/Ua=0.2$). }                       \label{fig205}
\end{center}
\par\vspace*{-3mm}
\end{figure}

Examples of numerically calculated streamlines and equipotential lines are shown in Fig.\ref {fig204} and Fig.\ref{fig205}.
The horizontal axis and the vertical axis indicate $ x / a, y / a $, respectively, and the flat plate is located at $ x = 0, y / a> -1 $.
Spacings of streamlines and equipotential lines ($ \gDelta \gPsi, \gDelta \gPhi $) are both $0.2Ua$ for $x>0$ and $0.4Ua$ for $x<0$.
(This also applies to Fig.\ref {fig207}-\ref{fig210} in the following sections.)
It is understood that the boundary condition on the flat plate is satisfied.
The broken line shows the wave height (the value of $ 1/10 $).
It is the characteristic of the flow that there is a strong circulating flow (reverse flow) around the flat plate.
A branch point of the flow (or a stagnation point) is recognized obliquely downward in front of the flat plate.
This flow approximates the flow when a strong circulation is added to the flow around a circular cross section in the infinite flow.
It is mainly due to that the linear solution deviates from its assumption near the intersection of the flat plate and the water surface.

\jsubsection{Weak singular solution}

In the previous section, the equation(\ref {equ210}) was used as the boundary condition of the reduced function.
In this section we shall find a reduced function that can take any value as $ \gPsi_H $ of the equation (\ref{equ207}).

This reduced function shall satisfy the equation (\ref{equ206}) and the following normalized condition.
\begin{equation} 
         \mathtt{Re}\{f_h(z) \} =1\ \ \ on\ \ \ x=0,\ -a<y<0.   \label{equ235}
\end{equation}

\begin{figure}[tp]
\vspace*{1mm}
\begin{center}
\includegraphics[width=0.90\linewidth,angle=0,bb=1 1 360 220,clip]{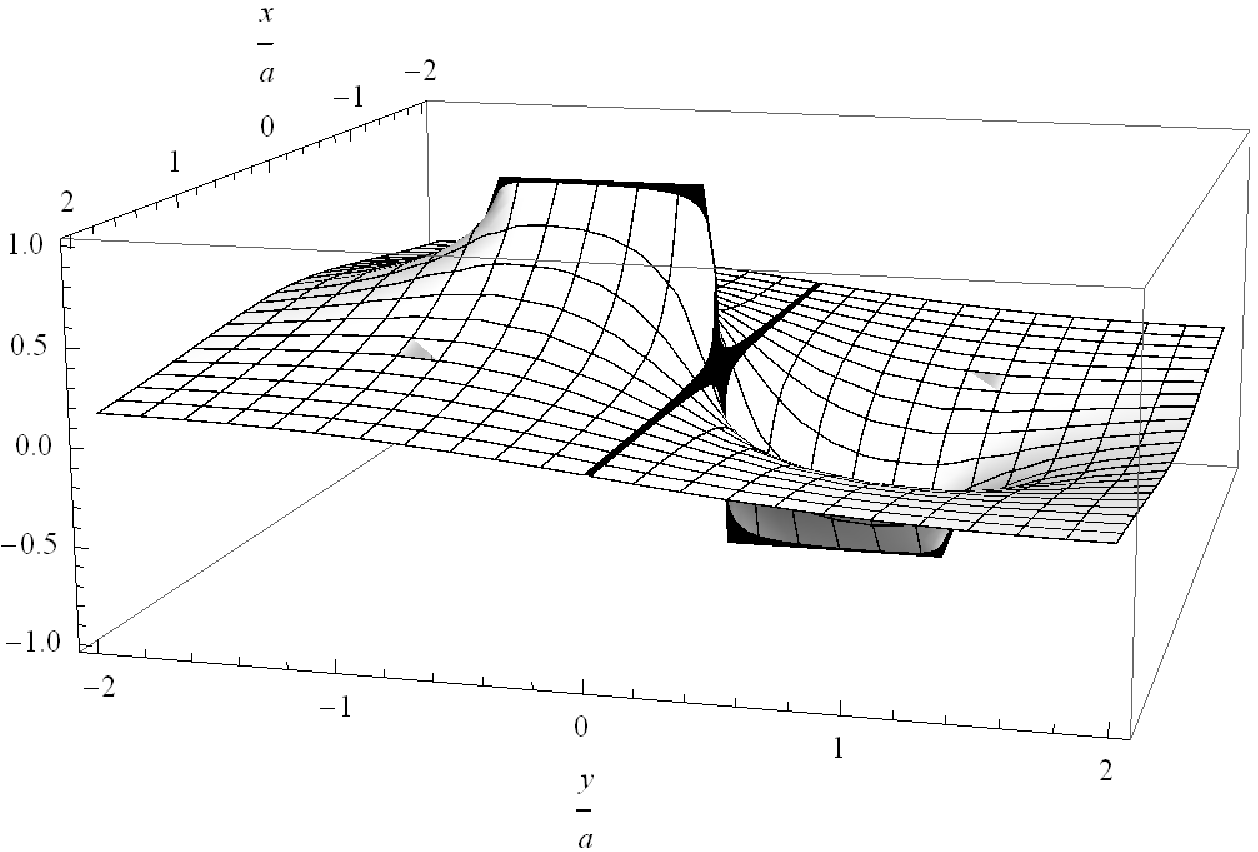}
\caption{3-D plot of $\mathtt{Re}\{f_h(z)\}$. }                         \label{fig206}
\end{center}
\par\vspace*{-3mm}
\end{figure}

\begin{figure}[thp]
\vspace*{1mm}
\begin{center}
\includegraphics[width=0.65\linewidth,angle=90,bb=352 285 1489 2100,clip]{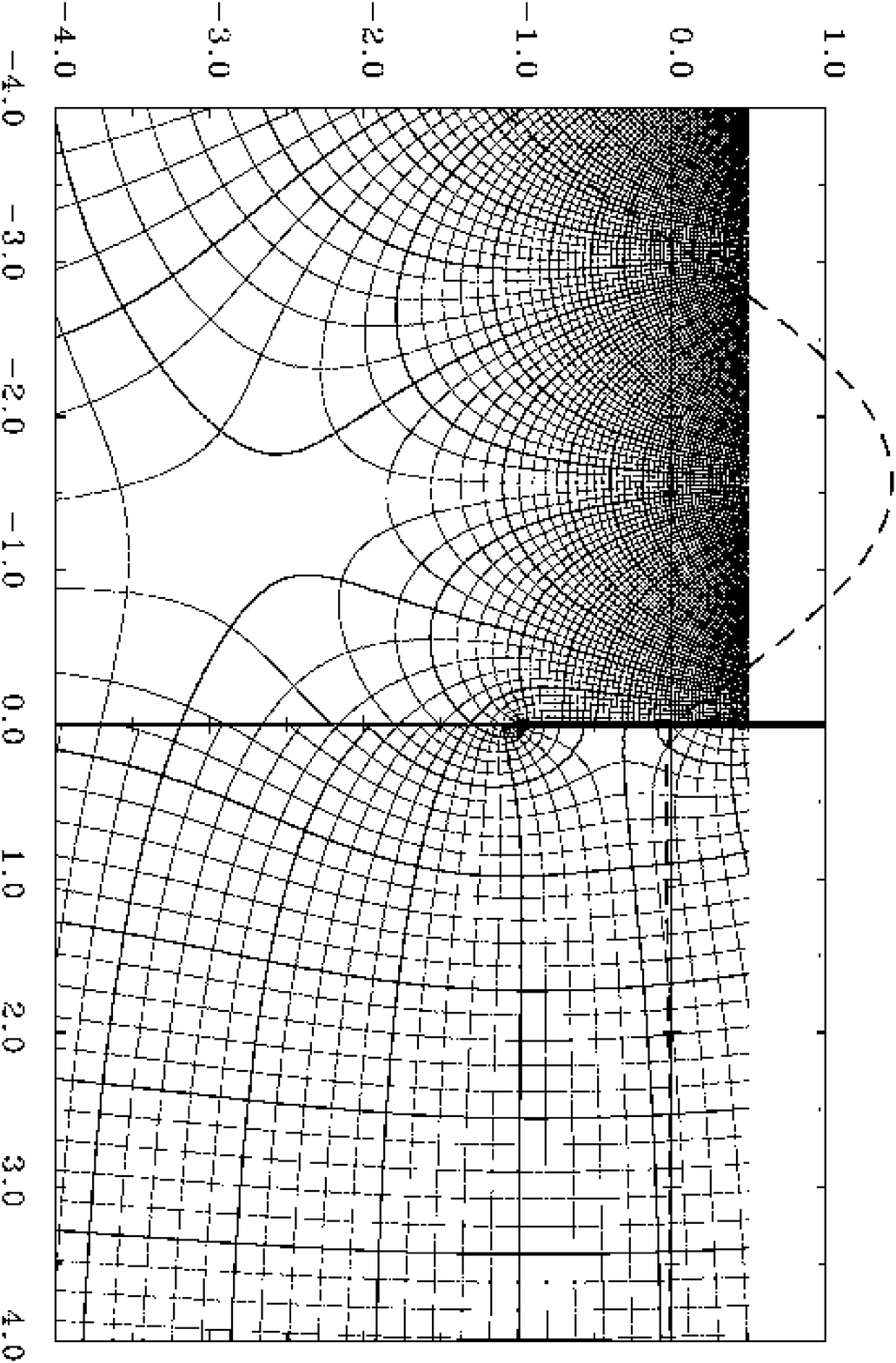}
\caption{Zero-vertical-flux flow ($\nu a=1.0,\ \gPsi_H/Ua=0.0,\ \gDelta\gPsi/Ua=0.2$). }                                                                                                        \label{fig207}
\end{center}
\par\vspace*{-3mm}
\vspace*{1mm}
\begin{center}
\includegraphics[width=0.65\linewidth,angle=90,bb=352 285 1489 2100,clip]{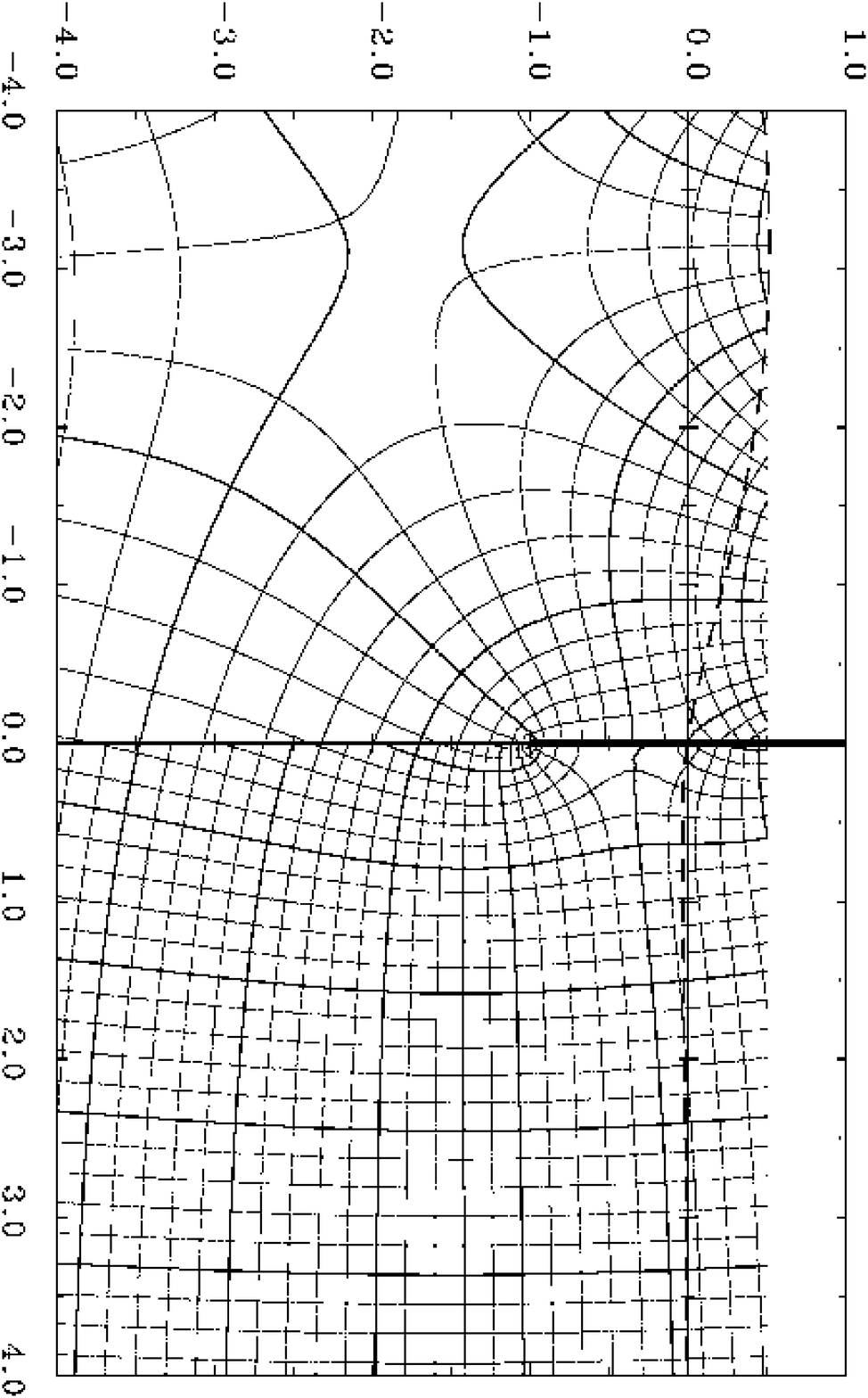}
\caption{Zero-vertical-flux flow ($\nu a=0.5,\ \gPsi_H/Ua=0.0,\ \gDelta\gPsi/Ua=0.2$). }                                                                                                        \label{fig207'}
\end{center}
\par\vspace*{-3mm}
\end{figure}

\begin{figure}[thp]
\vspace*{1mm}
\begin{center}
\includegraphics[width=0.65\linewidth,angle=90,bb=352 285 1489 2100,clip]{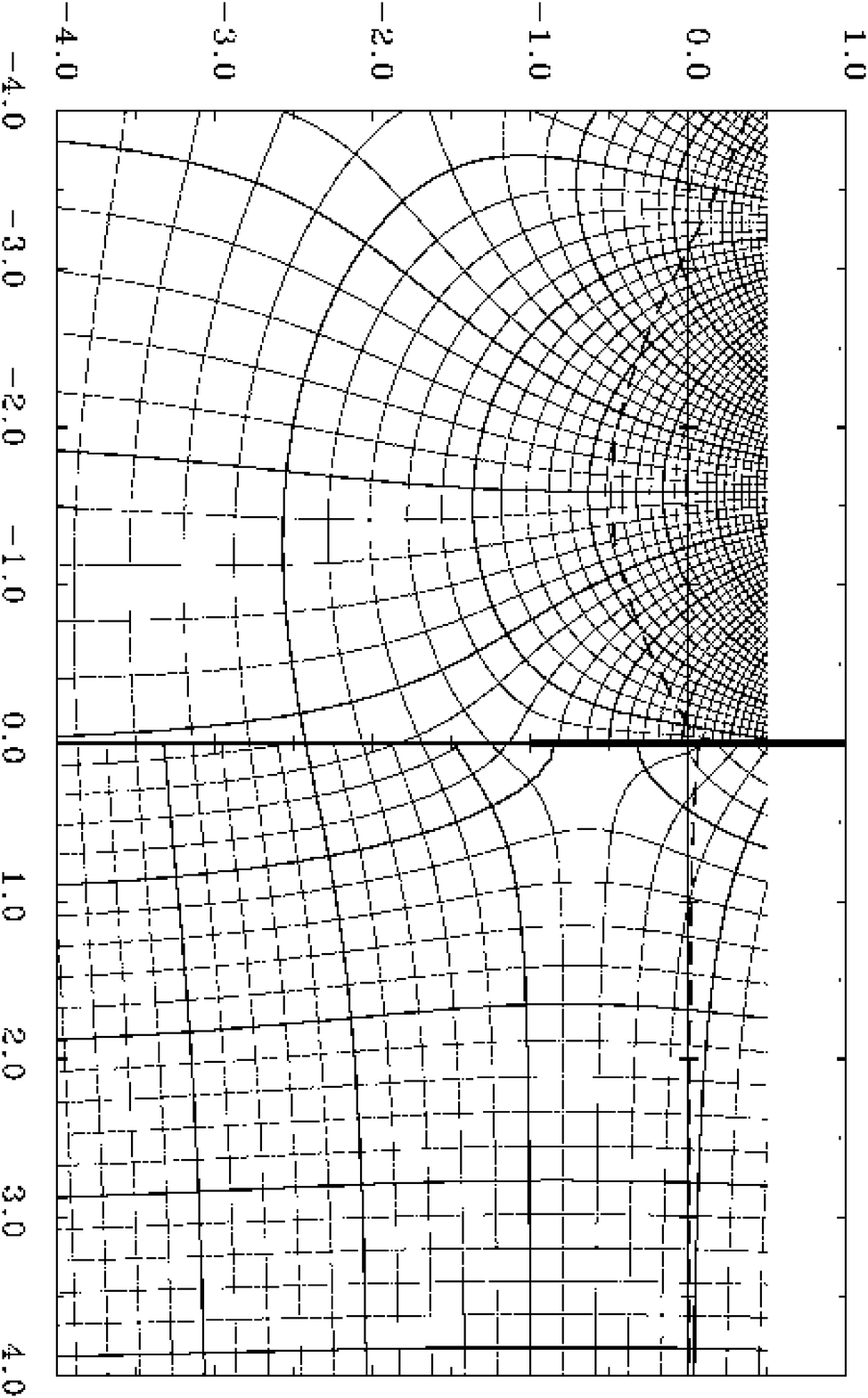}
\caption{Flow satisfying Kutta condition ($\nu a=1.0,\ \gPsi_H/Ua=0.664,\ \gDelta\gPsi/Ua=0.2$). }                                                                      \label{fig207''}
\end{center}
\par\vspace*{-3mm}
\end{figure}

\begin{figure}[thp]
\vspace*{1mm}
\begin{center}
\includegraphics[width=0.52\linewidth,angle=90,bb=499 311 1391 2078,clip]{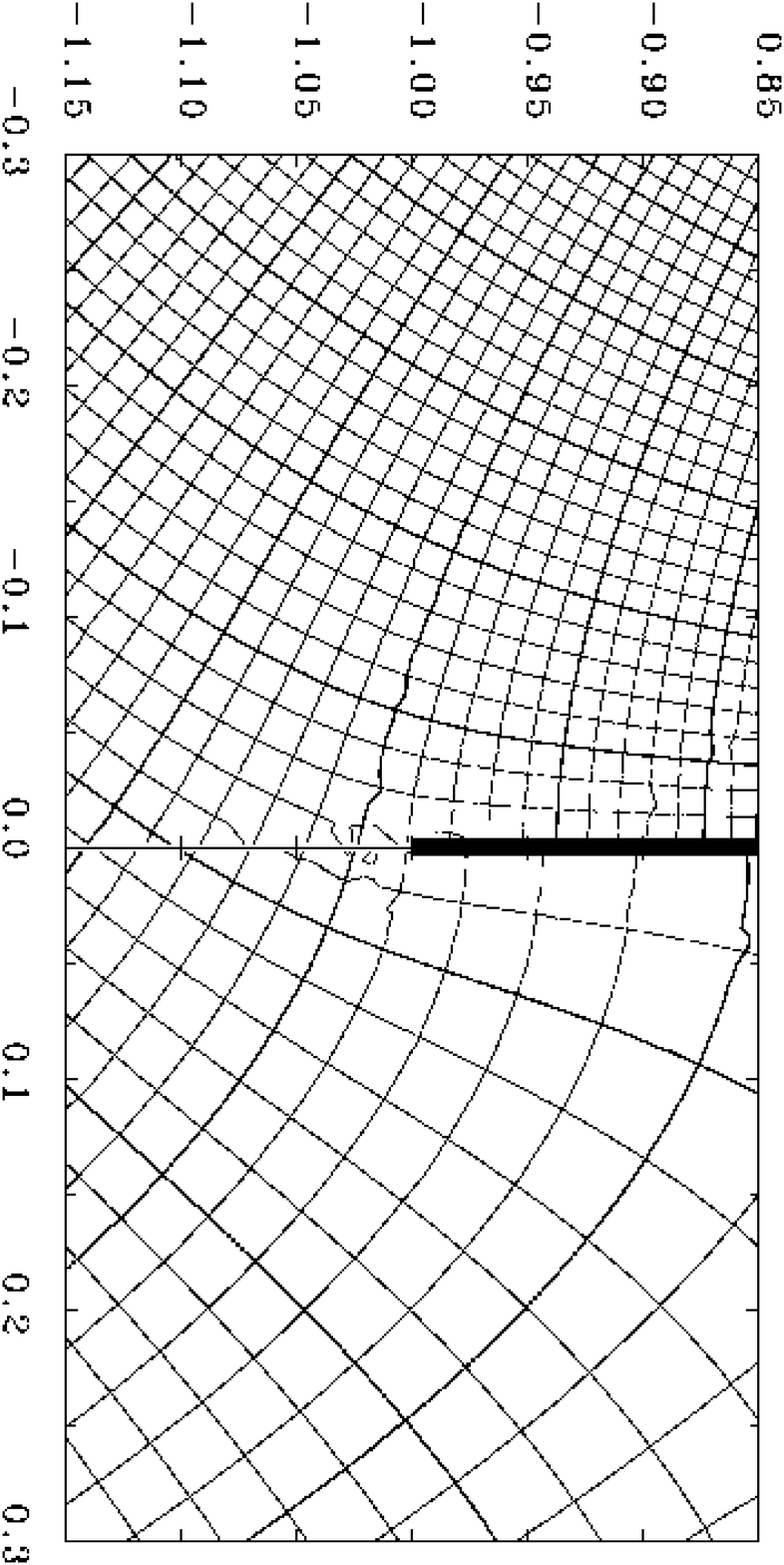}
\caption{Flow satisfying Kutta condition near the edge ($\nu a=1.0,\ \gPsi_H/Ua=0.664,\ \gDelta\gPsi/Ua=0.02$). }                                                       \label{fig208}
\end{center}
\par\vspace*{-3mm}
\end{figure}

\begin{figure}[thp]
\vspace*{1mm}
\begin{center}
\includegraphics[width=0.80\linewidth,angle=90,bb=414 712 1432 1784,clip]{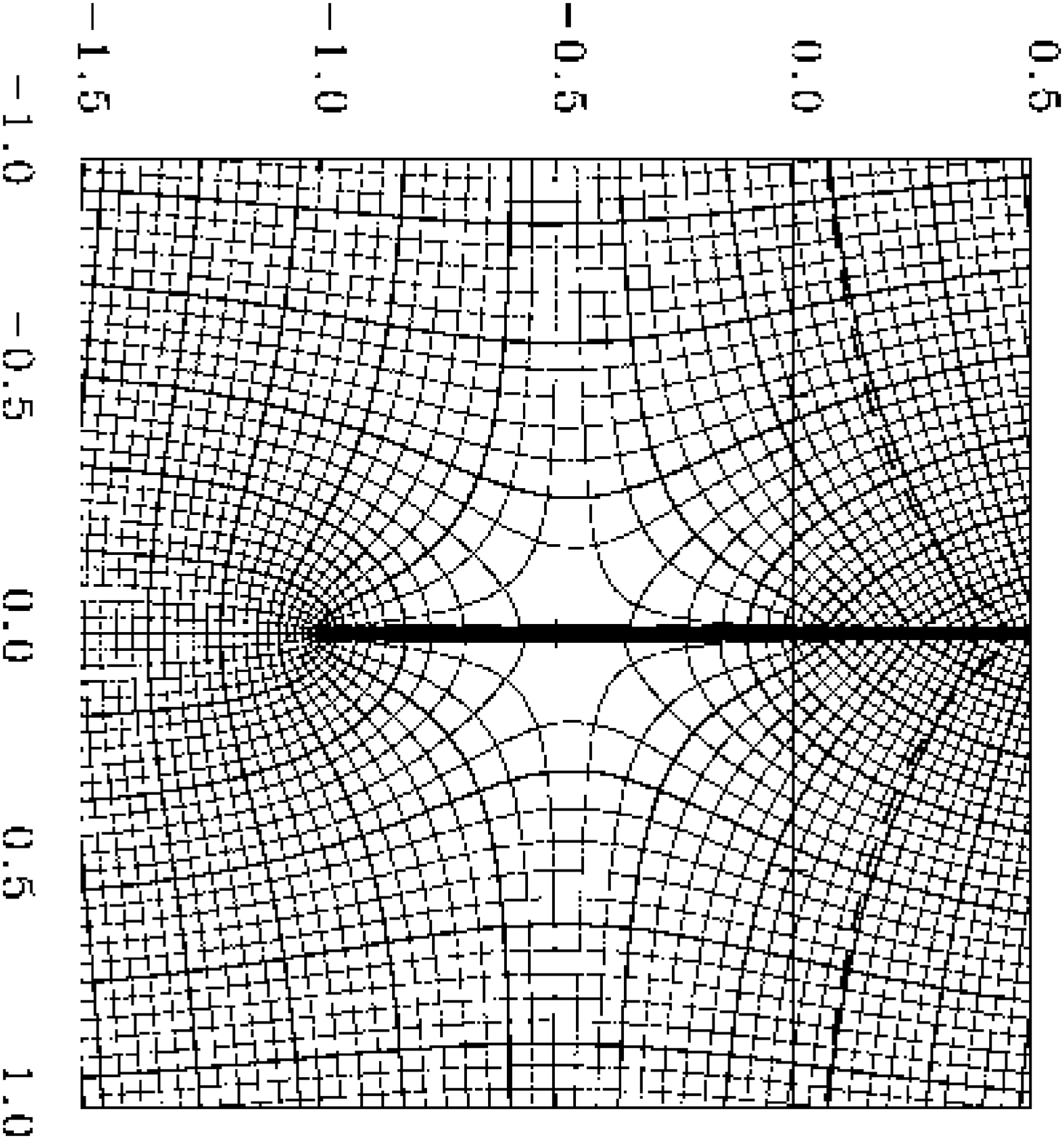}
\caption{Wave-free flow ($\nu a=1.0,\ \gPsi_H/Ua=0.494,\ \gDelta\gPsi/Ua=0.04$). }                                                                                                              \label{fig210}
\end{center}
\par\vspace*{-3mm}
\end{figure}

In the problem dealt with in this thesis, the solution is uniquely determined only for the regular solution got in the previous section. 
However, it has been shown that there is a special solution if it allows the flow velocity to have a logarithmic singularity (weak singularity) at the intersection of the object and the water surface\cite {U2}.
Therefore, the following is adopted as a reduced function satisfying the condition (\ref {equ235}).
\begin{equation} 
        f_h(z)=\frac{2}{\pi}\,i\log\Bigl[\sqrt{1+\frac{a^2}{z^2}} -\frac{a}{z}\Bigr].                                                                                                                           \label{equ236}
\end{equation}
The above expression has the following singularity at $ z = 0 $.
\begin{equation} 
        f_h(z)\sim \frac{2}{\pi}\,i\log\frac{z}{a}.                     \label{equ237}
\end{equation}
A three-dimensional plot of the real part of the reduced function $ f_h (z) $ is shown in Fig.\ref {fig206}.
It becomes $ 0 $ on the $ x $ axis, and it becomes the following on the $ y $ axis.
\begin{equation} 
        \mathtt{Re}\{f_h(z)\}=
                \begin{cases}
                        -1 \ \ \ for\ \ \ \ \ \ \ 0<y<a,                \\
                \ \ \,1\ \ \ for\ \ \  -a<y<0.
                \end{cases}                                                                             \label{equ238}
\end{equation}
The expression (\ref {equ236}) can be written with the following integral.
\begin{equation} 
         f_h(z)=-\frac{2}{\pi}\,i\int_z^\infty\frac{a}{z^2\sqrt{1+a^2/z^2}}dz.                                                                                                                                          \label{equ239}
\end{equation}
Since the integrand of this equation can be expressed by Laplace transformation, it can be expressed as follows by integration.
\begin{align} 
        f_h(z)=&-ai\int_z^\infty dz\int_0^\infty k\,J\hspace{-0.6mm}\mbf{H}(ak)e^{\mp kz}dk  \notag \\
        =&\mp ai\int_0^\infty J\hspace{-0.6mm}\mbf{H}(ak)e^{\mp kz}dk.                                                                                                                                                          \label{equ240}
\end{align}
The $ J \hspace {-0.6mm} \mbf {H} $ function in the above expression is abbreviation of the following expression.
\begin{equation} 
         J\hspace{-0.6mm}\mbf{H}(ak)=J_1(ak)\mbf{H}_0(ak)+J_0(ak)\bigl[\frac{2}{\pi}-\mbf{H}_1(ak)\bigr].                                                                               \label{equ241}
\end{equation}
Here, $ \mbf {H}_{0,1} $ and $ \mbf {L}_{0,1} $ appearing below are Struve functions.

The inverse transformation of the reduced function is set as the following and it is set as the sum of the local wave component and the free wave component.
\begin{align} 
        w_{h_0}(z)=&\frac{1}{L}\bigl[Uf_h(z) \bigr] \notag \\
        =&w_{h_0L}(z)+w_{h_0F}(z).                                                      \label{equ242}
\end{align}
When the same operation as in the previous section is performed, the following is obtained.
\begin{equation} 
        w_{h_0L}(z)=iaU\int_0^\infty\frac{1}{k\mp i\nu}J\hspace{-0.6mm}\mbf{H}(ak)e^{\mp kz}dk,                                                                                                 \label{equ243}
\end{equation}
\begin{equation} 
        w_{h_0F}(z)=
                \begin{cases}
                        \ \ 0\hspace{26mm}\ \ &for\ \ \ x>0,       \\
                        -2\pi aUI\hspace{-0.6mm}\mbf{L}(\nu a)\,e^{-i\nu z} &for\ \ \  x<0,
                \end{cases}                                                                     \label{equ244}
\end{equation}
where
\begin{align} 
        I\hspace{-0.6mm}\mbf{L}(\nu a)=&J\hspace{-0.6mm}\mbf{H}(-i\,\nu a)  \notag \\
                            =&-I_1(\nu a)\mbf{L}_0(\nu a)+I_0(\nu a)\bigl[\frac{2}{\pi}+\mbf{L}_1(\nu a)\bigr].                                                           \label{equ245}
\end{align}

The solution thus obtained does not satisfy the boundary condition on the flat plate.
\begin{equation} 
        \mathtt{Im}\{w_h(iy)\}=Const..                                                  \label{equ246}
\end{equation}
As with the $ W_1 (z) $ function in the previous section, we need to add a homogeneous solution $ w_ 0 (z) $.
\begin{equation} 
        w_h(z)=w_{h_0}(z)+C_2\,w_0(z).                                                   \label{equ247}
\end{equation}
The imaginary part of $ w_ {h_ 0} (z) $ is of the following form in the range of $ x = 0, -a \leqq y \leqq a $.
\begin{equation}
        \left.\begin{aligned}
                \frac{1}{Ua}\mathtt{Im}\{w_{h_0}(z)\}=&\bigl(T_{h_0}+\frac{1}{\nu a}\bigr)e^{\nu y}-\frac{1}{\nu a},                            \\
                T_{h_0}=&\frac{1}{Ua}\mathtt{Im}\{w_{h_0}(0)\}. \\
        \end{aligned}\hspace{0.5cm} \right\}                                    \label{equ248}
\end{equation}
Also, since the imaginary part of $w_0(z) $ is an expression(\ref{equ231}), it is as follows.
\begin{equation} 
         C_2=\frac{1}{K_0(\nu a)}\bigl(T_{h_0}+\frac{1}{\nu a}\bigr).                                                                                                                                                   \label{equ250}
\end{equation}
At this time it certainly fulfills the following boundary condition on the flat plate.
\begin{equation} 
        \mathtt{Im}\{w_h(iy)\}=-U/\nu\ \ \ for\ \ -a\leqq y\leqq 0.                                                                                                                                                     \label{equ251}
\end{equation}
Since the analytic value of the constant $T_{h_0}$ is unknown, it is necessary to obtain it numerically.

The velocity of the solution $ w_{h_0} (z) $ obtained here has a logarithmically weak singularity at the origin.
The function multiplied by a constant is also a solution.
We call this solution a weak singular (homogeneous) solution.

The function which is the sum of this weakly singular solution and the regular solution of the previous section is written as follows.
\begin{equation} 
        w(z)=w_R(z)+Cw_h(z).                                                                    \label{equ252}
\end{equation}
Then, the value of the stream function of this function on the flat plate becomes the following expression from the equations (\ref {equ209}) and (\ref{equ251}).
\begin{equation} 
         \gPsi_H=\frac{U}{\nu}\,(1-C).                                                   \label{equ253}
\end{equation}
Conversely, if we want to give $ \ gPsi_H $ you can do as follows.
\begin{equation} 
         C=1-\frac{\nu}{U}\,\gPsi_H.                                                    \label{equ254}
\end{equation}
The free waveform is as follows.
\begin{equation} 
         \eta_F(x)=A\sin\nu x.  \notag
\end{equation}
Then we have
\begin{equation} 
        A/a=A_R/a-2\pi C[I\hspace{-0.6mm}\mbf{L}(\nu a)+C_2I_0(\nu a)].                                                                                                                                                          \label{equ255}
\end{equation}
The wave resistance is written as follows.
\begin{align} 
        C_W=&\frac{R_W}{\rho ga^2}  \notag \\
        =&\frac{1}{4}(A/a)^2.                                                                   \label{equ256}
\end{align}

Let's first set $ C = 1 $.
Then, $\gPsi_H=0$.
Let's call this flow zero-vertical-flux flow because it means that the flow rate flowing out of the system from the water surface ($ y = 0 $) in the upstream of the flat plate becomes $ 0 $.
Examples of streamlines of this flow are shown in Fig.\ref {fig207} and Fig.\ref {fig207'}.
It is understood that the streamline passing through the infinite upstream water surface reaches the flat plate surface.
There is no strong circulating flow around the flat plate seen in regular solution.
From the flow of analytic continuation to $ y> 0 $ near the origin, we can see the weak singularity mentioned above.
The wave height indicated by the broken line is $ 0 $ at the origin.

Next, set the value of $ C $ as follows
\begin{equation} 
                C=-\nu a\frac{C_1}{C_2}.                                                        \label{equ256'}
\end{equation}
From the expression (\ref {equ252}), the term of $ w_0 (z) $ (accordingly, the flow singularity going around the lower end of the flat plate) disappears and the solution satisfying the Kutta condition at the lower end is obtained.
At this time, the value of the stream function on the flat plate becomes the following and the boundary condition is satisfied.
\begin{equation} 
        \mathtt{Im}\{ w(iy)\}=U(\frac{1}{\nu}+y)+Ua\frac{C_1}{C_2}.                                                                                                                                                     \label{equ256''}
\end{equation}
The streamline of this flow is shown in Fig.\ref {fig207''}.
The streamlines near the lower end of the flat plate of this flow are shown in Fig.\ref {fig208}.
No flow that goes around the lower end can be seen.
Some irregularities in streamlines and equipotential lines observed near the lower part of the flat plate is due to the fact that the accuracy of numerical integration is not good in the vicinity of $ x = 0 $.

\begin{figure}[thp]
\vspace*{1mm}
\begin{center}
\includegraphics[width=1.05\linewidth,angle=90,bb=60 223 569 671,clip]{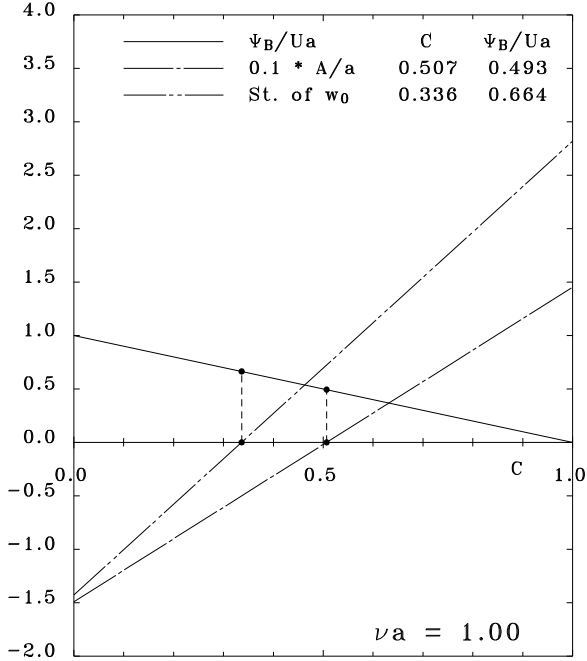}
\caption{$\gPsi_H$ et al. vs. C ($\nu a=1.0$). }                        \label{fig210'}
\end{center}
\par\vspace*{-3mm}
\end{figure}

We can also make a solution that the free wave height(\ref {equ255}) is 0.
The streamline of this wave-free flow is shown in Fig.\ref{fig210}.
The flow is totally symmetrical with respect to y-axis.

For convenience of understanding, the relation in the equation (\ref {equ254}) is shown in Fig.\ref{fig210'}.
The horizontal axis shows $ C $ (strength of weak singular solution), and the solid line shows the value of $\gPsi_H$.
The one-dot chain line shows the amplitude ($ 1/10 $, signed) of the free wave amplitude(\ref {equ255}).
The point of cutting off on the horizontal axis shows the state of wave-free.
The two-dot chain line shows the strength of the $ w_0 (z) $ function and the point of cutting off on the horizontal axis shows Kutta condition.

?@\\

\jsection{Discussion on the solution of Bessho-Mizuno(1962)\cite{BM}}

Translation of this section is omitted.

\jsection{Conclusions}

As a result of examining the representation of the complex potential for the flow around the semi-submerged flat plate in the two-dimensional uniform flow, the following conclusions were obtained.

\begin{enumerate}

\item As for the regular solution the following form of analytical representation was obtained for the perturbed complex potential.
\begin{equation} 
        w_R(z)=-\frac{1}{2}\nu a[w_2(z)+C'_1w_0(z)].  \notag 
\end{equation}
\item We showed that there is a homogenous solution $ w_h (z) $ with weak singularity at the intersection of the flat plate and still water surface (x=y=0) and obtained its analytical representation.

\item By adding homogeneous solutions multiplied by arbitrary constants to the regular solution, various flows can be made and analytical formulas of the wave making resistance of the flows are obtained.
\item It is highly likely that the analytical solution obtained by Bessho-Mizuno (1962)\cite{BM} does not satisfy exactly the boundary condition on the flat plate.

\end{enumerate}

\shaji{Acknowledgment} 

We would like to thank Dr. Hiroshi Isshiki who recommended us to translate this paper and submit to an international journal or arXiv.

 
\begin{furoku}


List the integral formulas used in the text. Some refer to Bateman \cite{B}. $a, \nu> 0 $.
\ \\
\begin{equation} 
        \int_0^\infty\frac{\nu}{k^2+\nu^2}J_0(ak)dk=\frac{\pi}{2}[I_0(\nu a)-\mbf{L}_0(\nu a)],                                                                                  \label{equA1}
\end{equation}
\begin{equation} 
        \int_0^\infty\frac{\nu}{k^2+\nu^2}J_2(ak)dk=\frac{1}{3}\nu a-\frac{\pi}{2}[I_2(\nu a)-\mbf{L}_2(\nu a)],                                                        \label{equA2}
\end{equation}
\begin{equation} 
        \int_0^\infty\frac{\nu\sin ak}{k^2+\nu^2}J_0(ak)dk=-K_0(\nu a)\sinh(\nu a),                                                                                                             \label{equA3}
\end{equation}
\begin{equation} 
        \int_0^\infty\frac{\nu\sin ak}{k^2+\nu^2}J_2(ak)dk=-\frac{2}{\nu a}+K_2(\nu a)\sinh(\nu a).                                                                             \label{equA4}
\end{equation}
$y\leqq -a$
\begin{equation} 
        \int_0^\infty\frac{\nu\cos ky}{k^2+\nu^2}J_0(ak)dk=\frac{\pi}{2}I_0(\nu a)\,e^{\nu y},                                                                                          \label{equA5}
\end{equation}
\begin{equation} 
        \int_0^\infty\frac{\nu\cos ky}{k^2+\nu^2}J_2(ak)dk=-\frac{\pi}{2}I_2(\nu a)\,e^{\nu y}.                                                                                          \label{equA6}
\end{equation}
$-a\leqq y\leqq 0$
\begin{equation} 
        \int_0^\infty\frac{k\cos ky}{k^2+\nu^2}J_0(ak)dk=K_0(\nu a)\cosh(\nu y),                                                                                                                                   \label{equA7}
\end{equation}
\begin{equation} 
        \int_0^\infty\frac{k\cos ky}{k^2+\nu^2}J_2(ak)dk=\frac{2}{(\nu a)^2}-K_2(\nu a)\cosh(\nu y),                                                                             \label{equA8}
\end{equation}
\begin{equation} 
        \int_0^\infty\frac{\nu\sin ky}{k^2+\nu^2}J_0(ak)dk=K_0(\nu a)\sinh(\nu y),                                                                                                                       \label{equA9}
\end{equation}
\begin{equation} 
        \int_0^\infty\frac{\nu\sin ky}{k^2+\nu^2}J_2(ak)dk=\frac{2}{\nu a^2}\,y-K_2(\nu a)\sinh(\nu y).                                                                          \label{equA10}
\end{equation}
$y\leqq -a$
\begin{equation} 
        \int_0^\infty\frac{k\sin ky}{k^2+\nu^2}J_0(ak)dk=-\frac{\pi}{2}I_0(\nu a)\,e^{\nu y},                                                                                                   \label{equA11}
\end{equation}
\begin{equation} 
        \int_0^\infty\frac{k\sin ky}{k^2+\nu^2}J_2(ak)dk=\frac{\pi}{2}I_2(\nu a)\,e^{\nu y}.                                                                                            \label{equA12}
\end{equation}

\end{furoku}
\lastpagecontrol{000mm}

\lastpagesettings

\end{document}

%% file: jasnaoe-e.tex
\makeatletter
\mark{{}{}}
\def\ps@headings{
  \let\@mkboth=\markboth
  \def\@evenfoot{}
  \def\@oddfoot{}
  \def\@evenhead{}
  \def\@oddhead{}}
\def\ps@myheadings{
  \let\@mkboth=\@gobbletwo
  \def\@evenfoot{}
  \def\@oddfoot{}
  \def\@evenhead{
   \hbox to \textwidth
   {\small {\thepage}\hspace{15pt}\hfil \leftmark}}
  \def\@oddhead{
   \hbox to \textwidth
   {\small \hfil  \rightmark \hfil\hspace{15pt}\thepage}}
  \def\chaptermark##1{}
  \def\sectionmark##1{}
  \def\subsectionmark##1{}}
\def\footnoterule{
  \kern -3\p@ \hrule width 1.0\columnwidth \kern 2.6\p@}
\long\def\@makefntext#1{
  \par\hangindent 20pt \noindent
  \hbox to 1.8em{\hss\@thefnmark}\hspace{5pt}#1}

\def\maketitle{\par
 \begingroup
 \def\thefootnote{*\arabic{footnote}}
 \def\@makefnmark{\hbox to 5pt{$^{\@thefnmark}$\hss}}
 \if@twocolumn \twocolumn[\@maketitle]
 \else \newpage \global\@topnum\z@ \@maketitle
 \fi
 \thispagestyle{headings}
 \@thanks
 \endgroup
 \setcounter{footnote}{0}
 \let\maketitle\relax \let\@maketitle\relax
 \gdef\@thanks{}\gdef\@author{}\gdef\@title{}%
 \let\thanks\relax}
\def\@maketitle{
\newpage
 \null
 \vskip 1.0em
\begin{center} {\LARGE \@title \par}\end{center}
\vskip 1.3em

\@ifundefined{@eauthor}{}{
{\small \@eauthor}
\gdef\@eauthor{}}
\vskip 2.5em

\@ifundefined{@abst}{}{
{\scriptsize
\begin{center}{\footnotesize\bf Summary}\vspace*{-.5em}\end{center}
\begin{quotation} \@abst \end{quotation}}
\gdef\@abst{}}
\vskip 3.0em
 
}
\long\def\eauthor#1{\long\gdef\@eauthor{#1}}
\long\def\abst#1{\long\gdef\@abst{#1}}

\def\thesection{\arabic{section}.}

\def\section{\@startsection {section}{1}{\z@}
{4.0ex plus 1.0ex minus .2ex}{2.0ex plus .1ex}{\small\bf}}
\def\subsection{\@startsection{subsection}{2}{\z@}
{2.5ex plus 0.5ex minus .2ex}{0.5ex plus .1ex}{\footnotesize\bf}}
\def\subsubsection{\@startsection{subsubsection}{3}{\z@}
{1.0ex plus 0.5ex minus .2ex}{0.5ex plus .1ex}{\footnotesize\bf}}
\def\paragraph{\@startsection{paragraph}{4}{\z@}
{1ex plus .3ex minus .2ex}{-1em}{\small\bf}}
\def\subparagraph{\@startsection{subparagraph}{4}
{\parindent}{1ex plus .3ex minus .2ex}{-1em}{\footnotesize\bf}}

\newcommand{\jsection}[1]{
\setcounter{subsection}{0}
\setcounter{subsubsection}{0}
\begin{center}
\addtocounter{section}{1}
\vspace{1.75pt}
{\small\bf\baselineskip=12pt \thesection\hspace*{2.5mm} #1}
\vspace{1.25pt}
\end{center}
}
\def\jsubsection#1{\subsection{\hspace*{-1.0mm}#1}}

\newcommand{\shaji}[1]{
\begin{center}
\vspace{1.75pt}
{\small\bf\baselineskip=12pt #1}
\vspace{1.25pt}
\end{center}
}

\def\eqnarray{%
 \stepcounter{equation}%
 \let\@currentlabel=\theequation
 \global\@eqnswtrue
 \global\@eqcnt\z@
 \tabskip\@centering
 \let\\=\@eqncr
 $$\halign to \displaywidth\bgroup\@eqnsel\hskip\@centering
 $\displaystyle\tabskip\z@{##}$&\global\@eqcnt\@ne
 \hfil$\displaystyle{{}##{}}$\hfil
 &\global\@eqcnt\tw@$\displaystyle\tabskip\z@{##}$\hfil
 \tabskip\@centering&\llap{##}\tabskip\z@\cr}
\def\leftcapmargin{40pt}
\def\rightcapmargin{15pt}
\long\def\@makecaption#1#2{%
 \vskip 10\p@ \footnotesize\baselineskip=4.0mm
 \@tempdimb=\leftcapmargin
 \@tempdima\hsize\advance\@tempdima-\@tempdimb
 \@tempdimb=\rightcapmargin
 \advance\@tempdima-\@tempdimb
 \newbox\tempbox
 \setbox\tempbox=\hbox{#1}
 \setbox\@tempboxa=\hbox{\hskip\leftcapmargin #2\hskip\rightcapmargin}
 \ifdim \wd\@tempboxa >\hsize
  \hbox to\hsize{\hfil \hskip-\the\wd\tempbox \hskip\leftcapmargin #1 
   \parbox[t]\@tempdima{#2}\hskip\rightcapmargin}%
  \else \hbox to\hsize{\hfil #1 #2\hfil}\fi
}
 \def\fnum@figure{Fig.$\;$\thefigure$\;\;$}
 \def\fnum@table{Table$\;$\thetable$\;\;$}
\def\@cite#1#2{$^{\hbox{\tiny{#1\if@tempswa , #2\fi})}}$}
\def\thebibliography#1{%
\begin{center}
\section*{References}
\end{center}
\list
{\arabic{enumi})}{\settowidth\labelwidth{#1)}\leftmargin\labelwidth
 \advance\leftmargin\labelsep
 \usecounter{enumi}}
 \def\newblock{\hskip .11em plus .33em minus .07em}
 \sloppy
 \sfcode`\.=1000\relax}

\let\oriixpt\ixpt
\def\ixpt{%
  \oriixpt%
\def\boldmath{%
  \@ifundefined%
    {ninmib}
    {
     \global\font\ninmib=cmmib9
     \global\font\ninsyb=cmsy9
     \global\font\ninlyb=lasyb10 at 9pt\relax
     \@addfontinfo\@ixpt%
     {\def\boldmath{%
       \everymath{\mit}%
       \everydisplay{\mit}
       \@prtct\@nomathbold
       \textfont\@ne\ninmib
       \textfont\tw@\ninsyb
       \textfont\lyfam\ninlyb
       \@prtct\@boldtrue}}}
     {}%
     \@ixpt\boldmath}%
}
\let\oriviiipt\viiipt
\def\viiipt{%
  \oriviiipt%
\def\boldmath{%
  \@ifundefined%
    {egtmib}
    {
     \global\font\egtmib=cmmib8
     \global\font\egtsyb=cmsy8
     \global\font\egtlyb=lasyb10 at 8pt\relax
     \@addfontinfo\@viiipt%
     {\def\boldmath{%
       \everymath{\mit}%
       \everydisplay{\mit}
       \@prtct\@nomathbold
       \textfont\@ne\egtmib
       \textfont\tw@\egtsyb
       \textfont\lyfam\egtlyb
       \@prtct\@boldtrue}}}
     {}%
     \@viiipt\boldmath}%
}
\def\sloppy{\tolerance=9999 \hfuzz=.5\p@ \vfuzz=.5\p@}

\begingroup
 \catcode`\|=0 \catcode`\\=13 |gdef|verbh@@k{|catcode`|\=13 |let\=|yen}
|endgroup

\begingroup \catcode`|=0 \catcode`[=1 \catcode`]=2 \catcode`\{=12
 \catcode`\}=12 \catcode`\\=13
 |gdef|@xverbatim#1\end{verbatim}[#1|end[verbatim]]
 |gdef|@sxverbatim#1\end{verbatim*}[#1|end[verbatim*]]
|endgroup

\def\verbatim{\@verbatim \frenchspacing\@vobeyspaces \verbh@@k \@xverbatim}

\@namedef{verbatim*}{\@verbatim \verbh@@k \@sxverbatim}

\@ifundefined{LaTeXe}{%
\def\verb{\begingroup \catcode``=13 \@noligs
 \tt \let\do\@makeother \dospecials \verbh@@k \@ifstar{\@sverb}{\@verb}}
}{%
\def\verb{\relax\ifmmode\hbox\else\leavevmode\null\fi
  \bgroup \verbatim@font\@noligs
    \verb@eol@error \let\do\@makeother \dospecials \verbh@@k
    \@ifstar\@sverb\@verb}
}
\newif\if@lastpagecolumnalign \@lastpagecolumnaligntrue
\if@lastpagecolumnalign 
\newdimen\lastp@geheight \lastp@geheight=10mm
\newsavebox{\lastp@gebox}
\long\def\endabstract#1{\gdef\@endabstract{#1}} 
\def\lastpagecontrol{\@ifnextchar [{\l@stpagecontrol}%
 {\l@stpagecontrol[\z@]}}
\def\l@stpagecontrol[#1]#2{\global\lastp@geheight=#2%
 \@ifundefined{maxsize}{}{\global\advance\maxsize-#2}
 \@ifundefined{@endabstract}{}{
 \global\sbox{\lastp@gebox}{%
  \begin{minipage}{\textwidth}\vspace*{#1}%
   \hrule width \textwidth \vspace{1ex}%
   \begin{list}{}{\setlength{\leftmargin}{0.15\textwidth}%
    \listparindent=10pt \parsep=0pt
    \topsep=0pt \partopsep=0pt
   \setlength{\rightmargin}{\leftmargin}\small}%
     \item \ignorespaces\hspace*{\listparindent}\ignorespaces
                 \@endabstract
   \end{list}%
   \vspace{1ex} 
  \end{minipage}}}%
  \@tempdima\ht\lastp@gebox \advance\@tempdima\dp\lastp@gebox
 \ifdim\@tempdima>\lastp@geheight
  \@tempdima\lastp@geheight \global\lastp@geheight=0pt
 \else
  \global\advance\lastp@geheight -\@tempdima
  \@tempdima\lastp@geheight \global\lastp@geheight\textheight
 \fi
  \def\footnoterule{\null}
  \insert\footins{\footnotesize
  \interlinepenalty\interfootnotelinepenalty
  \splittopskip\footnotesep
  \splitmaxdepth \dp\strutbox \floatingpenalty \@MM
  \hsize\textwidth \@parboxrestore
   \ifdim\lastp@geheight=\z@\else\usebox{\lastp@gebox}\fi%
  \vspace*{\@tempdima}}}
\def\lastpagesettings{\@ifnextchar [{\l@stpagesettings}%
 {\l@stpagesettings[\z@]}}
\def\l@stpagesettings[#1]{%
 \ifdim\lastp@geheight=\z@
  \onecolumn\null\vspace*{#1}\noindent\usebox{\lastp@gebox}%
 \fi}
\fi 
\makeatother

%% file: deffile-e.tex
\newlength{\mybaselineskip}
\newlength{\myleftmargin}
\newlength{\mytopmargin}
\newcommand{\bdm}{\begin{displaymath}}
\newcommand{\edm}{\end{displaymath}}
\newcommand{\be}{\begin{equation}}
\newcommand{\ee}{\end{equation}}
\newcommand{\bea}{\begin{eqnarray}}
\newcommand{\eea}{\end{eqnarray}}
\newcommand{\bc}{\begin{center}}
\newcommand{\ec}{\end{center}}
\newcommand{\bfl}{\begin{flushleft}}
\newcommand{\efl}{\end{flushleft}}
\newcommand{\bfr}{\begin{flushright}}
\newcommand{\efr}{\end{flushright}}
 
\newcommand{\mbf}[1]{\mbox{\boldmath$#1$}}

\newcommand{\gPhi}{\mathnormal{\Phi}}
\newcommand{\gPsi}{\mathnormal{\Psi}}

\newcommand{\gDelta}{\mathnormal{\Delta}}

\newcommand{\ename}[2]{#1 \ {\it #2}}

\newenvironment{furoku}{%
\setcounter{equation}{0}%
\setcounter{section}{0}%
\setcounter{subsection}{0}%
\setcounter{figure}{0}%
\setcounter{table}{0}%
\def\theequation{A\arabic{equation}}
\def\thesection{A\arabic{section}.}

\def\thefigure{A\arabic{figure}}
\def\thetable{A\arabic{table}}
\par\medskip%
\begin{center}{\small\bf Appendix}\end{center}
\vspace*{-1mm}}{ }